\documentclass[useAMS,usenatbib]{mn2e}

\usepackage[utf8]{inputenc}
\usepackage{amsmath}
\usepackage{amsfonts}
\usepackage{amssymb}
\usepackage[english]{babel}
\usepackage{tabu}

\usepackage{graphicx}

\usepackage{mathtools}

\newcommand{\eg}{{\rm e.g.}}

\usepackage{setspace}
 
\usepackage[switch]{lineno}

\def\H0{{\rm ~km~s^{-1}~Mpc^{-1}}}

\def\cf{{\rm c.f.~\/}}
\def\la{\mathrel{\hbox{\rlap{\hbox{\lower4pt\hbox{$\sim$}}}{\raise2pt\hbox{$<$}}}}}
\def\ga{\mathrel{\hbox{\rlap{\hbox{\lower4pt\hbox{$\sim$}}}{\raise2pt\hbox{$>$}}}}}
\def\d25{D$_{25}$}

\title[Subphotospheric Dissipation]{Confronting GRB prompt emission with a model for subphotospheric dissipation}
\author[B. Ahlgren et al.]{Bj\"{o}rn Ahlgren$^{1}$\thanks{E-mail:
bjornah@kth.se}, Josefin Larsson$^{1}$, Tanja Nymark$^{1}$, Felix Ryde$^{1}$, Asaf Pe'er$^{2}$
\\
$^{1}$KTH Royal Institute of Technology, Departement of Physics, and the Oscar Klein Centre, AlbaNova, SE-106 91 Stockholm, Sweden \\
$^{2}$Physics Department, University College Cork, Cork, Ireland}
\begin{document}
\date{2015 August}

\maketitle

\label{firstpage}

\begin{abstract}
The origin of the prompt emission in gamma-ray bursts (GRBs) is still an unsolved problem and several different mechanisms have been suggested. Here we fit Fermi GRB data with a photospheric emission model which includes dissipation of the jet kinetic energy below the photosphere. The resulting spectra are dominated by Comptonization and contain no significant contribution from synchrotron radiation. In order to fit to the data we span a physically motivated part of the model's parameter space and create DREAM (\textit{Dissipation with Radiative Emission as A table Model}), a table model for {\scriptsize XSPEC}. We show that this model can describe different kinds of GRB spectra, including GRB~090618, representing a typical Band function spectrum, and GRB~100724B, illustrating a double peaked spectrum, previously fitted with a Band+blackbody model, suggesting they originate from a similar scenario. We suggest that the main difference between these two types of bursts is the optical depth at the dissipation site.
\end{abstract}


\section{Introduction}
For decades, the prompt $\gamma$-ray spectra of gamma-ray bursts (GRBs) have been fitted with the empirical Band function \citep{1993ApJ...413..281B}. Although often producing good fits to GRB spectra, the model does not represent any physical scenario. In order to extract physical information from the observations we need to introduce physically motivated models and fit them to data. 

There are several reasons why such models should include emission from the jet photosphere: (\textit{i}) Synchrotron radiation fails to explain the observed GRB spectra due to the line of death \citep{Preeceetal_1998A_ApJ} and spectral width \citep{2015MNRAS.447.3150A}; (\textit{ii}) Some GRBs have spectra which are close to pure Planck functions \citep{2004ApJ...614..827R, Ghirlandaetal_2013A_MNRAS, 2015ApJ...800L..34L}; (\textit{iii}) Many GRBs are well described by models comprising a blackbody (BB) and an additional component \citep{Rydeetal_2010A_ApJ, 2011ApJ...727L..33G, 2012ApJ...757L..31A, Burgess_2014A_ApJ}. %2013MNRAS.433.2739I , 2013ApJ...770...32G

At the same time, most bursts produce spectra not consistent with the simplest version of photospheric emission; in general the low-energy slope is far too soft to be accounted for by a BB. However, it has been realised that we should not expect a pure Planck function from the photosphere in a relativistic outflow, since there are natural mechanisms which broaden the spectrum, \eg~subphotospheric dissipation \citep{2005ApJ...628..847R, Pe'eretal_2006A_ApJ,2006A&A...457..763G,2015arXiv150203055C}, or geometric broadening \citep{2008ApJ...682..463P,Lundmanetal_2013A_MNRAS}. In this work we consider a model based on the former and fit it to GRB data from the \textit{Fermi Gamma-ray Space Telescope}.  Subphotospheric dissipation has already been suggested as the emission mechanism in some bursts, \eg ~ for GRB~090902B \citep{2011MNRAS.415.3693R,2012MNRAS.420..468P} and GRB~110920 \citep{2015arXiv150305926I}, but a full physical model has not previously been fitted to data. The analysis presented in this Letter provides a first proof-of-concept of fitting such a model to data.

Throughout this paper we are assuming standard $\Lambda$-CDM cosmology, a Hubble constant of $H_0 = 69.6$ and a cosmological constant, $\Omega_{\Lambda} = 0.714$. All errors are 1 sigma unless otherwise stated.

\section{The Model}
\label{themodel}
%\subsection{The physical scenario}
The physical scenario we consider is based on the fireball model and the radiation is treated using the code described in \cite{2005ApJ...628..857P}. For a review of the fireball model, see \eg~\cite{2006RPPh...69.2259M}. In our picture a progenitor releases a luminosity $L_{0,52} = L_{0} 10^{-52}$ erg s$^{-1}$ (not to be confused with the observed luminosity), in a relativistic, collimated and magnetised jet of electrons, baryons, and photons. The jet is accelerated up to the saturation radius $r_{\mathrm{s}}$ such that $r_{\mathrm{s}} \sim r_\mathrm{0} \eta$, where the bulk flow Lorentz factor $\Gamma = \eta$ and where $\eta = L_{0,52}/\dot{M}c^2$ is the dimensionless entropy and $r_\mathrm{0}$ the nozzle radius. We assume dissipation to occur at a radius $r_{\mathrm{d}}$, defined by the corresponding optical depth $\tau$. A fraction $\varepsilon_{\mathrm{d}} L_{0,52}$ of the energy is dissipated by some, in principle unspecified, process, \eg ~internal shocks \citep{2005ApJ...628..847R}, magnetic reconnection \citep{1994MNRAS.270..480T, 2005A&A...430....1G} or hadronic collision shocks \citep{2010MNRAS.407.1033B}. Here we assume $r_0 = r_{\mathrm{d}} \Gamma^{-2}$, thus using internal shocks for this mechanism \citep{1999MNRAS.306L..39M}. In practice, the only effect of this assumption is that it sets the initial photon temperature, $T_{0}(r_0)$. The dissipated energy is divided between magnetic fields, which receive a fraction $\varepsilon_{\mathrm{b}}\varepsilon_{\mathrm{d}} L_{0,52}$, and the electrons, receiving $\varepsilon_{\mathrm{e}}\varepsilon_{\mathrm{d}} L_{0,52}$. A fraction $\varepsilon_{\mathrm{pl}}$ of the electrons take on a power law distribution and a fraction $(1-\varepsilon_{\mathrm{pl}})$ assume a Maxwellian distribution.

Photons and particles interact via Compton and inverse Compton scattering, pair production/annihilation and synchrotron self-absorption, and the electrons emit synchrotron radiation. The code follows the spectral evolution of the electrons and photons over one dynamical time, $t_{\mathrm{dyn}} = r_{\mathrm{d}}/c$ with a fine time resolution. Although the code does not simulate the hydrodynamical evolution of the jet, we can evaluate the time evolution in a GRB by performing time-resolved spectroscopy and assuming that the jet properties are driven by changes of the central engine. From the dissipation radius to the photospheric radius there should be adiabatic expansion, effectively cooling the jet, however this effect is neglected since it is small and comparable to other uncertainties \citep{2004ApJ...613..448P}. 

\begin{figure}
\begin{center}
\includegraphics[scale=0.33, trim=1cm 0cm 0cm 1.4cm]{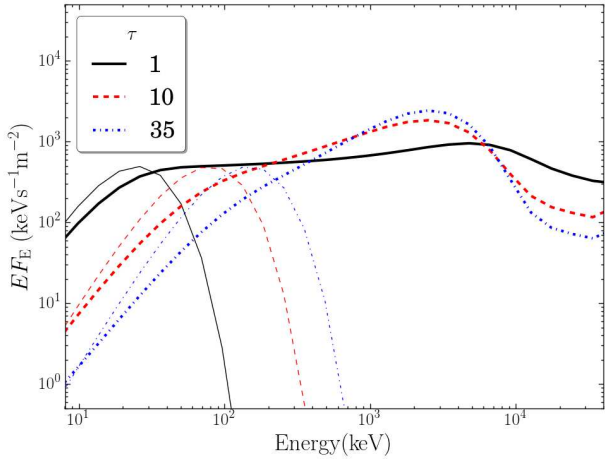}
\caption{Spectra obtained from the code for different values of the $\tau$ of the dissipation site. The other parameters are $\Gamma = 250,L_{0,52}= 10, \varepsilon_{\mathrm{pl}} = 0, \varepsilon_{\mathrm{b}} = 10^{-6}, \varepsilon_{\mathrm{e}} = 0.9$ and $\varepsilon_{\mathrm{d}} = 0.2 $. In addition for each spectrum we also show the BB spectrum of the seed photons, plotted with the same style and colour as the spectrum, but with thinner lines.}
\label{code_output1}
\end{center}
\end{figure}

In order to be able to fit this model to data we create a grid of models by running the code for different input parameters. The grid is then turned into a table model for {\scriptsize XSPEC}. We will refer to the model as DREAM (\textit{Dissipation with Radiative Emission as A table Model}). This way we may perform relatively fast fits even though the simulations are computationally expensive. For this study we have chosen to confine ourself to a four-dimensional parameter space, using the parameters $\tau , \Gamma ,L_{0,52}, \varepsilon_{\mathrm{d}}$, keeping the other three parameters fixed at $\varepsilon_{\mathrm{b}} = 10^{-6}, ~ \varepsilon_{\mathrm{e}} = 0.9$ and $\varepsilon_{\mathrm{pl}} = 0$. This choice reflects a scenario where the vast majority of the energy goes to the electrons, which take on a completely Maxwellian distribution as they are heated \citep{2012ApJ...756..174L}, and where we have weak magnetic fields, yielding negligible synchrotron radiation. This allows us to address the question of whether the observed spectra can be explained without this process. The values for the parameters used are $\tau = 1,5,10,20,35, ~ \Gamma = 50,100,250,500, ~ L_{0,52} = 0.1, 1, 10, 100, 300, ~ \varepsilon_{\mathrm{d}} = 0.1, 0.2, 0.3, 0.4, 0.5 $, yielding a table model consisting of 500 simulations and spanning a physically motivated part of the parameter space. In {\scriptsize XSPEC} the model obtains two additional parameters; a redshift, $z$ and an additional normalization which is proportional to the observed photon flux. In order to eliminate degeneracies we keep these two parameters constant for each burst. Additionally, to make sure that the resulting fits are not strongly affected by the step size in the table we also created a much finer grid spanning a smaller part of the parameter space. A comparison with the original model showed no significant differences.

The shape of the resulting spectra for different input parameters has been discussed by \cite{Pe'eretal_2006A_ApJ}. Here we summarise the main points relevant in the context of our spectral fits. Note that the effects described below are non-linear and that the effect of changing one parameter partially depends on the values of the other parameters. The code produces an output spectrum in terms of photon emissivity against photon energy in the comoving jet frame. In Fig.~\ref{code_output1} we have plotted $E F_E$ spectra in observer frame for varying values of $\tau$. We note how the shape of the spectrum changes with increasing $\tau$, becoming more thermalized, ultimately approaching a Wien spectrum for higher values of $\tau$, as expected. Furthermore, we have in Fig.~\ref{code_output1} included the initial BBs, corresponding to each spectrum's thermal seed photons. The figure illustrates that a low-energy spectral slope softer than Rayleigh-Jeans can be obtained as the thermal seed-photons are up-scattered. This effect is stronger for high optical depths. 

The luminosity, $L_{0,52}$, corresponds to the amount of energy we have in the spectrum as well as the comoving proton number density, since $L_{0,52} \propto n_{\mathrm{p}}$ \citep{Pe'eretal_2006A_ApJ}, and thus a higher $L_{0,52}$ corresponds to a higher normalisation. Considering $r_{\mathrm{d}} \propto L_{0,52}/\tau \Gamma^3 $ \citep{Pe'eretal_2006A_ApJ}, along with the photon temperature going as $T = T_0 (r_{\mathrm{d}}/r_{\mathrm{s}})^{-2/3}$ \citep{2005ApJ...628..847R}, we note that a higher $L_{0,52}$ also moves the initial BB towards lower energies. In contrast, considering the dependence on $\Gamma$ in the expression above, an increase in $\Gamma$ results in fewer but more energetic photons, hence shifting the spectrum to higher energies and lowering the normalization. Another important effect of high $\Gamma$ is pair production. When pairs are created in sufficient numbers they increase the effective optical depth, resulting in a considerably stronger thermalization. Lastly, $\varepsilon_{\mathrm{d}}$ increases the temperature of the electrons, with the normalised temperature $\theta \propto \varepsilon_{\mathrm{d}}$ \citep{Pe'eretal_2006A_ApJ}, yielding an increasingly higher second peak in the spectrum due to Comptonisation. Hence, as $\varepsilon_{\mathrm{d}}$ approaches zero, the spectrum tends to the initial photon Planck function.

Due to assumptions made in the code, we expect a general, underlying uncertainty in the parameter values that produce a given spectrum, estimated to amount up to a factor of $\sim$2. The main contributions to this uncertainty is that the code is 1 dimensional (and thus spatial effects are not considered) and that it does not account for the hydrodynamical evolution of the jet (see also \citealt{2004ApJ...613..448P,2005ApJ...628..857P}). It should be noted that this uncertainty is not captured by the statistics when performing fits.

\section{Observations and Fitting the Model}
\label{fitting_section}
We illustrate the application of the DREAM model by fitting it to two different, bright \textit{Fermi} GRBs; GRB~090618 and GRB~100724B. While GRB~090618 is well fitted by a Band function, GRB~100724B was the first example of a burst with a significant additional BB component \citep{2011ApJ...727L..33G}. GRB~090618 is analysed using Gamma-ray Burst Monitor (GBM) data \citep{2009ApJ...702..791M} from the NaI and BGO detectors. For GRB~100724B, we used GBM data from the NaI and BGO detectors as well as LAT-LLE data. The LLE data are a type of LAT data designed for the use with bright transients to bridge the energy range of the LAT and GBM \citep{2010arXiv1002.2617P}. For both bursts we selected NaI detectors seeing the GRB at an off-axis angle lower than 60 degrees and the BGO detector as being the best aligned of the two BGO detectors. The spectra were fitted in the energy ranges 8--1000~keV (NaI), 200--40~000~keV (BGO) and 30--1000~MeV (LAT-LLE). For a more extensive analysis of the high energy data for GRB~100724B, see \cite{2013ApJS..209...11A} who find evidence for an exponential cut-off at high energies.

Our spectral analysis is time resolved, using a signal-to-noise-ratio (SNR) = 40 in order to bin the data. As a consistency check we have also performed the binning with Bayesian Blocks, and the results were not changed considerably. The analysis is performed using HEASARC's {\scriptsize XSPEC} 12.8.1g, with pgstat statistics and with PHA (Pulse Height Analyser) data.

\subsection{GRB~090618}
\begin{figure*}
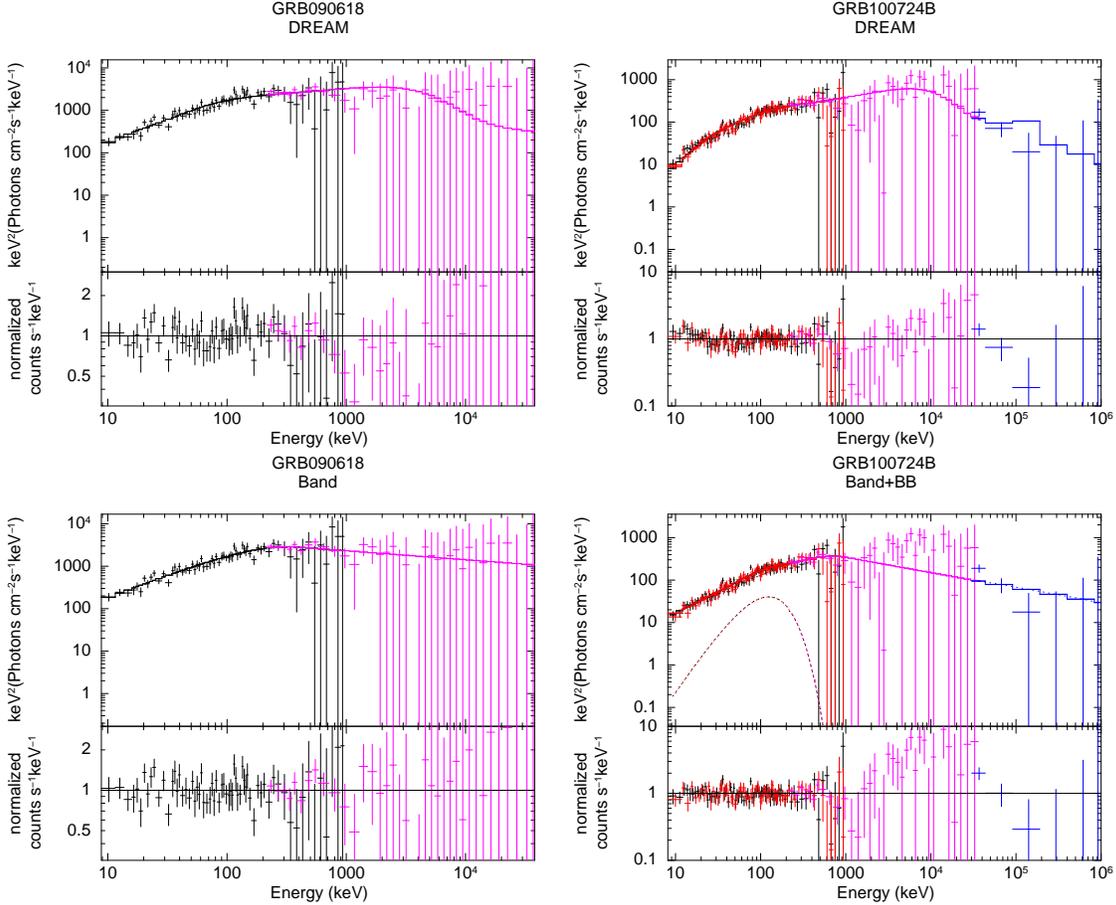

\begin{center}
\includegraphics[scale=0.3,angle=270]{grb090618_DREAM.eps}
\includegraphics[scale=0.3,angle=270]{grb100724_DREAM.eps}
\includegraphics[scale=0.3,angle=270]{grb090618_band.eps}
\includegraphics[scale=0.3,angle=270]{grb100724_band+bb.eps}
\caption{Fit results for GRB~090618 and GRB~100724B at $65.3-65.7$~s and $25.8-33.5$~s, respectively. The first column shows GRB~090618; fitted with the DREAM model above and with the Band function below. Data from NaI 4 and BGO 0 are shown in black and magenta, respectively. The right column shows fits of GRB~100724B with the DREAM model above and with Band+BB below. Data from NaI 0,1 and BGO 0 are shown in black, red and magenta respectively. LAT-LLE data are shown in blue. The lower panels in these plots show the ratio between the model and the data points. Data have been re-binned for visual clarity, to errors of 4 sigma.} %(intervals indicated in Fig.~\ref{Lightcurves})
\label{allfits}
\end{center}
\end{figure*}

GRB~090618 has a fluence of $2.684 \times 10^{-4} \pm 4 \times 10^{-7}$~erg~cm$^{-2}$ \citep{FermiGBMcat_2014} and a redshift $z = 0.54$ \citep{2009GCN..9518....1C}, which translates into an observed luminosity $L = 2.8 \times 10^{51} $~erg~s$^{-1}$. It is a long GRB with 90 per cent of the flux emitted during $T_{90} = 112.39$~s. The fit results for the first five time bins are presented in Table~\ref{bestfit_tm_grb090618}, whereas the full table is available online. In Fig.~\ref{allfits}, we present the bin at $65.3-65.7$~s as an example. The fit with the DREAM model yields $\tau=17.1^{+1.5}_{-1.4},~\Gamma=239^{+2}_{-2},~L_{0,52}=33.0^{+1.5}_{-1.5}~\varepsilon_{\mathrm{d}}=0.100^{+0.001} $ and pgstat/degrees of freedom (d.o.f.) $= 251/240$. This test statistic shows that the DREAM model provides an acceptable fit to the data. For the same bin, also in Fig.~\ref{allfits}, we present the corresponding fit with a Band function, with the best-fitting parameters $\alpha = -0.92^{+0.08}_{-0.07}, ~\beta = -2.33^{+0.12}_{-0.16}, ~ E_{\mathrm{peak}} = 340^{+42}_{-42}$~keV and the test statistic pgstat/d.o.f. $= 236/240$ \footnote{Note that these test statistics should not be used for model to model comparison as the models are not nested. Furthermore, such comparison is not meaningful since only the DREAM model represents a physical scenario.}.

\begin{table}
\caption{Best-fitting parameters for the first five time bins for GRB~090618 fitted with the DREAM model. $\varepsilon_{\mathrm{d}}$ is pegged at its lowest value. For the complete time evolution, the full table is available online.}
\label{bestfit_tm_grb090618}
\setlength\tabcolsep{2pt}
\begin{tabular}{lccccc} 
\hline \\[-9pt] Time (s) & pgstat/d.o.f. & $\tau$ & $\Gamma$ & $ L_{0,52}$ & $\varepsilon_{\mathrm{d}}$ \\ \hline \\[-7pt]
-2.0-4.0 & $306/240 $ & $14.0^{+1.8}_{-2.3}$ & $233^{+2}_{-2}$ & $4.51^{+0.38}_{-0.56}$ & $0.100^{+0.002}_{}$ \\
[2pt]4.0-8.0 & $347/240 $ & $5.43^{+0.69}_{-0.53}$ & $242^{+2}_{-3}$ & $6.06^{+0.46}_{-0.51}$ & $0.100^{+0.002}_{}$ \\
[2pt]8.0-11.7 & $294/240 $ & $5.02^{+0.54}_{-0.41}$ & $233^{+3}_{-3}$ & $5.14^{+0.29}_{-0.32}$ & $0.100^{+0.002}_{}$ \\
[2pt]11.7-15.2 & $247/240 $ & $4.30^{+0.71}_{-0.60}$ & $232^{+5}_{-6}$ & $5.00^{+0.30}_{-0.28}$ & $0.100^{+0.002}_{}$ \\
[2pt]15.2-19.5 & $304/240 $ & $2.47^{+0.32}_{-0.27}$ & $242^{+5}_{-5}$ & $3.54^{+0.16}_{-0.18}$ & $0.100^{+0.002}_{}$ \\
\hline
\end{tabular}
\end{table}

\subsection{GRB~100724B}
GRB~100724B has a fluence of $2.174 \times 10^{-4} \pm 5 \times 10^{-7}$~erg~cm$^{-2}$, $T_{90} = 114.69$~s \citep{FermiGBMcat_2014}, and an unknown redshift. In order to conserve four free parameters, we set the redshift to $1$ when performing the fits to this burst. The most prominent effect of varying $z$ for this burst is a change in $L_{0,52}$ and $\varepsilon_\mathrm{d}$ due to the change in flux.

The fit results for the first five time bins are presented in Table~\ref{bestfit_tm_grb100724}, whereas the full table is available online. In Fig.~\ref{allfits}, we show a fit with the DREAM model at $25.8-33.5$~s, after the trigger, as well as the corresponding fit with a Band+BB function. The DREAM model fit yields the parameters $\tau =4.91^{+1.25}_{-0.30},~\Gamma = 443^{+6}_{-9},~L_{0,52}= 41.9^{+2.0}_{-2.1}~\varepsilon_{\mathrm{d}} = 0.12^{+0.01}_{-0.01} $ with pgstat/d.o.f. $=406/383$ and the fit with Band+BB yields the parameters $\alpha = -1.06^{+0.05}_{-0.06},~\beta = -2.36^{+0.04}_{-0.08},~E_{\mathrm{peak}} = 712^{+149}_{-149}$~keV, $kT = 31.7^{+3.7}_{-3.9}$~keV with pgstat/d.o.f. $=401/381$.

\begin{table}
\caption{Best-fitting parameters for the first five time bins in GRB~100724B fitted with the DREAM model. For the complete time evolution, the full table is available online.}
\label{bestfit_tm_grb100724}
\setlength\tabcolsep{2pt}
\begin{tabular}{lccccc} %\multicolumn{1}{p{8pt}}{\centering pgstat/ \\ \centering d.o.f.} 
\hline \\[-9pt] Time (s) & pgstat/d.o.f. & $\tau$ & $\Gamma$ & $ L_{0,52}$ & $\varepsilon_{\mathrm{d}}$ \\ \hline \\[-7pt]
-1.0-9.4 & $526/383 $ & $17.5^{+1.5}_{-2.2}$ & $426^{+2}_{-11}$ & $69.8^{+4.1}_{-5.8}$ & $0.100^{+0.002}_{}$ \\
[2pt]9.4-12.1 & $454/383 $ & $9.90^{+0.76}_{-0.88}$ & $472^{+3}_{-5}$ & $160^{+4}_{-5}$ & $0.126^{+0.011}_{-0.006}$ \\
[2pt]12.1-15.0 & $406/383 $ & $12.3^{+1.1}_{-1.5}$ & $469^{+4}_{-5}$ & $163^{+4}_{-4}$ & $0.100^{+0.002}_{}$ \\
[2pt]15.0-17.1 & $481/383 $ & $9.26^{+0.96}_{-1.05}$ & $470^{+5}_{-5}$ & $180^{+9}_{-7}$ & $0.163^{+0.015}_{-0.014}$ \\
[2pt]17.1-19.4 & $434/383 $ & $9.89^{+0.57}_{-0.80}$ & $475^{+3}_{-5}$ & $182^{+4}_{-7}$ & $0.138^{+0.011}_{-0.008}$ \\
\hline
\end{tabular}
\end{table}

\section{Discussion and conclusion}
As shown in section~\ref{fitting_section}, we find that the DREAM model provides acceptable fits to the data. In addition, the parameters are determined with small statistical errors, in the range of $3\% - 23\% $ for all parameters. These errors are of the same magnitude as the errors in the Band- or Band+BB-parameters. However, these errors must be assessed keeping in mind the much more essential uncertainties due to assumptions in the model, as noted above. In the case of $\varepsilon_{\mathrm{d}}$ it reaches its lowest allowed value of our examined parameter space in most of the bins for these bursts. Hence, the corresponding one sided statistical errors, shown in Table~\ref{bestfit_tm_grb090618}, should be seen only as an indication of the fit's sensitivity to an increasing $\varepsilon_{\mathrm{d}}$. In our current configuration we find that no pairs of parameters are highly degenerate or correlated around the best-fitting values. Some degeneracies do exist, however, most notably between $\tau$ and $L_{0,52}$, $\varepsilon_\mathrm{d}$ and $L_{0,52}$ as well as $\Gamma$ and $\tau$. The pegged value of $\varepsilon_\mathrm{d}$ implies that the value of $L_{0,52}$ might in reality be higher in these bursts than what is found in our fits. However, a significantly lower value of $\varepsilon_{\mathrm{d}}$ is not expected since that would yield a spectrum close to a Planck function, see the discussion in section \ref{themodel}. We have examined progressively lower values of $\varepsilon_{\mathrm{d}}$ for the best-fitting parameter values of GRB~090618, which indicate that we should have $\varepsilon_{\mathrm{d}} > 0.01$, as the observed spectrum is not close to a Planck function. However, due to the non-linearity of the problem, it is difficult to give a robust lower limit $\varepsilon_{\mathrm{d}}$. 

When considering the fits in more detail we start by examining GRB~090618; this burst exhibits a typical Band function, in the sense that its best-fitting parameter values lie close to the average values of all GBM GRBs \citep{2014ApJS..211...12G}. With $\alpha \sim -0.9$ the soft part of the spectrum is certainly not what would normally be associated with a BB. This is of particular interest as it is often assumed that synchrotron emission is needed to obtain such values of $\alpha$, whereas our model in its current state has no synchrotron component. Within the DREAM model the soft low-energy slope is instead created as a result of Comptonization of the seed BB, \cf Fig~\ref{code_output1}. In addition, in the Band fits the value of $\alpha$ depends on the assumption that the spectrum has a single peak. If the true spectral shape is instead weakly double-peaked (as suggested by the fits with the DREAM model, see Fig.~\ref{allfits}) this approximation will push $\alpha$ towards softer values.

GRB~100724B, in contrast to GRB~090618, is well fitted by a Band+BB model and thus represents a different category of burst. However, when fitted with the DREAM model both bursts appear slightly double peaked, see Fig.~\ref{allfits}. This seems to suggest that these two bursts may originate from the same physical processes and should be considered to belong to the same category. The reason why adding a BB results in a greater statistical improvement for GRB~100724B than for GRB~090628 is likely driven by the distance between the two peaks, $E_2/E_1$, being significantly larger for GRB~100724B (qualitatively seen in Fig.~\ref{allfits}), with $E_1$ and $E_2$ corresponding to the low- and high-energy peaks, respectively. Furthermore, the temperature of the seed BB determines how much of the hard low-energy spectrum is seen in the GBM energy range, and thus whether a BB component can be identified (note that observed break in energy, $E_1$, does not correspond exactly to the seed photon energy), see Fig.~\ref{code_output1}. Within the DREAM model we find that, when $\varepsilon_{\mathrm{d}}$ is constant, the value of $E_2/E_1$ is primarily set by $\tau$, with a higher $\tau$ yielding a lower $E_2/E_1$. In this context we note that the optical depth may in reality be higher than what is given by $\tau$, due to pair production in the jet, which becomes increasingly important with high $\Gamma$. This has no effect on any of our conclusions in this paper. 

An important conclusion that can be drawn is that, assuming this model to be correct, the Band $E_{\mathrm{peak}}$ does not represent the true peak energy of the spectrum, but is an average value of the two peaks. This point is something one would need to consider when interpreting the $E_{\mathrm{peak}}-E_{\mathrm{iso}}$ \citep{2002A&A...390...81A} relation. Another feature of the model is the pair annihilation peak, seen at $10^{5}$~keV for GRB~100724B in Fig.~\ref{allfits}. Note that the placement of this peak is uncertain due to the unknown redshift. The DREAM model also predicts a sharp cut-off at higher energies, above the fitted energy range.

In the evolution of best-fitting parameters of these two bursts there is a general trend that $\Gamma$ varies little throughout the bursts, as one would expect from general fireball dynamics. The same is true for $\varepsilon_{\mathrm{d}}$, although we note that this parameter reaches its lowest allowed value in most of the bins, see discussion above. Furthermore the behaviour of $L_0$ tends to approximately follow the light curves, as expected from the nearly constant $\varepsilon_{\mathrm{d}}$. The evolution of $\tau$ tends approximately to that of $L_{0,52}$, resulting in a weakly varying dissipation radius, see section~\ref{themodel}. We note that the time scale for the parameter evolution is well beyond the time scale of the microphysical processes of the emission. As pointed out by \cite{2008ApJ...682..463P,2011ApJ...732...49P}, there is an intermediate time scale of light propagation in the scenario of subphotospheric dissipation, when taking \eg~geometric effects into account, which might be long enough to induce a temporal smearing between emission from multiple, discrete dissipation sites. This would imply a connection between smooth pulses in the lightcurve and subphotospheric dissipation. \cite{2011ApJ...737...68B} argues that this smearing is suppressed enough to allow for separate treatment of the individual pulses. In that case, as we assume that the dynamical conditions are roughly constant during one pulse, the observed parameter evolution reflects the changing dynamical properties of the central engine and/or dissipation process.

In summary, we have shown that the DREAM model provides fits of acceptable quality to the data, using GRB~090618 and GRB~100724B as examples. Notably these fits has been performed with negligible synchrotron radiation. We have shown that these two bursts' spectra are similar, having two spectral breaks as their main feature. We have thus demonstrated that they can be considered to originate from the same process, whereas previous fits to these bursts have claimed different components. The main difference between the two bursts is the distance between the peaks, which we chiefly attribute to the parameter $\tau$ in our current setting. In a future work we will extend the model to include additional free parameters, in particular to include the Synchrotron regime.

\section*{Acknowledgements}
This work was supported by the Göran Gustafsson Stiftelse and the Swedish National Space Board. Some of the simulations were performed using resources at PDC Centre for High Performance Computing (PDC-HPC). The \textit{Fermi}-LAT Collaboration acknowledges support for LAT development, operation and data analysis from NASA and DOE (United States), CEA/Irfu and IN2P3/CNRS (France), ASI and INFN (Italy), MEXT, KEK, and JAXA (Japan), and the K.A.~Wallenberg Foundation, the Swedish Research Council and the National Space Board (Sweden). Science analysis support in the operations phase from INAF (Italy) and CNES (France) is also gratefully acknowledged.
\bibliographystyle{mn2e_v2}
\bibliography{SUPER_bibliography}

\label{lastpage}
\end{document}